%% file: main.tex
\documentclass[10pt,conference]{IEEEtran} 
\usepackage[utf8]{inputenc}

\usepackage{amsfonts}
\usepackage{amsmath}
\usepackage{amssymb} 
\usepackage{amsbsy}
\usepackage{microtype}

\usepackage{wrapfig}
\usepackage{enumitem}
\usepackage{soul}
\usepackage{subfigure}
\usepackage{graphbox}
\usepackage[braket, qm]{qcircuit}
\usepackage{bbm}

\usepackage[percent]{overpic}

\usepackage{graphicx,calc}
\newlength\myheight
\newlength\mydepth
\settototalheight\myheight{Xygp}
\newcommand*\inlinegraphics[1]{%
  \settototalheight\myheight{Xygp}%
  \settodepth\mydepth{Xygp}%
  \raisebox{-\mydepth}{\includegraphics[height=\myheight]{#1}}%
}

\usepackage[hidelinks]{hyperref}
\usepackage{tikz}
\usepackage{xcolor} 

\input{colored_circles}

\def\equationautorefname~#1\null{Equation~(#1)\null}

\newcommand{\appurl}{https://graphstatevis.github.io/app}
\newcommand{\GraphStateVis}{\href{\appurl}{\textsc{GraphStateVis}}}

\title{\href{\appurl}{GraphStateVis}: Interactive Visual Analysis  \\ of Qubit Graph States and their Stabilizer Groups}
 \author{
     \IEEEauthorblockN{Matthias Miller\IEEEauthorrefmark{1} and  Daniel Miller\IEEEauthorrefmark{2}\IEEEauthorrefmark{4}
     }
     \IEEEauthorblockA{\IEEEauthorrefmark{1}Datenanalyse und Visualisierung, University of Konstanz,  Universit\"atsstr. 10,  DE-78465 Konstanz}
     \IEEEauthorblockA{\IEEEauthorrefmark{2}Institut f\"ur Theoretische Physik III, Heinrich-Heine-Universit\"at D\"usseldorf, 
   Universi\"atsstr. 1,   DE-40225   D\"usseldorf }
 }

\makeatletter 
 \hypersetup{pdftitle = {GraphStateVis: Interactive Visual Analysis of Qubit Graph States and their Stabilizer Groups},
	     pdfauthor = {Matthias Miller, Daniel Miller},
	     pdfsubject = {Quantum Technology},
	     pdfkeywords = {GraphStateVis, visual analytics, visualization, visualisation, interactive, tool, application, app, demo, prototype, construct, design, build, develop, graph,    adjacency matrix, quantum, qubit,  state, stabilizer, stabiliser, sector length, distribution, Pauli weight, SLD, noise threshold, entanglement, hardware efficient, shallow, circuit, variational, VQE, measurements, connectivity, subgraphs, entangled basis, separable basis, tensor product basis, TPB} 
	    }
\makeatother

\date{May 2021}

\newcommand{\diag}{\mathrm{diag}}
\newcommand{\CZ}{\textsc{\MakeLowercase{CZ}} } 
\newcommand{\FF}{\mathbb{F}}
\newcommand{\RR}{\mathbb{R}}
\newcommand{\CC}{\mathbb{C}}
\newcommand{\Tr}{\mathrm{Tr}}
\newcommand{\wt}{\mathrm{wt}}
\newcommand{\swt}{\mathrm{swt}} 

\newcommand{\Gcon}{G_\mathrm{con}}
\newcommand{\GHZn}{\ket{\mathrm{GHZ}^n}}
\newcommand{\psiTheta}{\ket{\Psi(\boldsymbol{\theta})}}

\newcommand{\ibmqSydney}{\textit{ibmq\_sydney}}

\newcommand{\zoomtofit}{\inlinegraphics{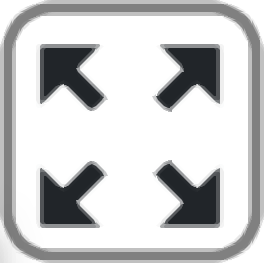}}
\newcommand{\forceUnLock}{\inlinegraphics{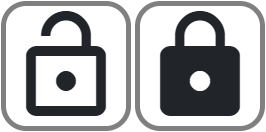}}
\newcommand{\copyGraphId}{\inlinegraphics{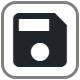}}
\newcommand{\copyUrlIcon}{\inlinegraphics{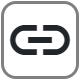}}
\newcommand{\parityColoringIcon}{\inlinegraphics{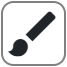}}
\newcommand{\distillationIcon}{\inlinegraphics{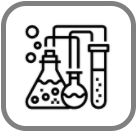}}

\input{figures}

\begin{document}

\maketitle

\begingroup\renewcommand\thefootnote{\textsection}
\footnotetext{Current address: IBM Quantum, IBM Research Europe – Zurich\\  E-Mail: {dmi(at)zurich.ibm.com}}
\endgroup

\input{0_abstract}

\input{1_Introduction}
\input{2_PrototypeDesign}
\input{3_UseCase}
\input{4_Conclusion} 
 
\bibliographystyle{IEEEtran}
\bibliography{libraryIEEE}

\end{document}

%% file: colored_circles.tex
\usepackage{pdfrender}

\usepackage[skins]{tcolorbox}
\newcommand{\boxSep}{0pt}
\newcommand{\boxRule}{0.5pt}
\newcommand{\boxTopLine}{2pt}
\newcommand{\boxBottomLine}{0pt}
\newcommand{\boxLeftLine}{1.2pt}
\newcommand{\boxRightLine}{1.2pt}
\newcommand{\boxArc}{2pt}

\newlength{\dpcircle}
\newlength{\rcircle}
\newlength{\dcircle}

\newcommand{\docircle}[4]{%
  \setlength{\dpcircle}{\dp\strutbox}%
  \setlength{\dcircle}{\dpcircle}%
  \addtolength{\dcircle}{\ht\strutbox}%
  \setlength{\rcircle}{0.5\dcircle}%
  \setlength{\unitlength}{1sp}%
  \begin{picture}(\number\dcircle,0)
    \color{#1}
    \put(\number\rcircle,\number\dpcircle){\circle*{\number\dcircle}}
    \color{#2}
    \put(\number\rcircle,\number\dpcircle){\circle{\number\dcircle}}
    \put(\number\rcircle,0){\makebox[0pt]{\textcolor{#3}{#4}}}
  \end{picture}%
}

\definecolor{pitch_c}{rgb}{0.976, 0.031, 0.349}
\definecolor{pitch_d}{rgb}{0.969, 0.576, 0.000}
\definecolor{pitch_e}{rgb}{0.969, 0.957, 0.035}
\definecolor{pitch_f}{rgb}{0.898, 0.031, 0.965}
\definecolor{pitch_g}{rgb}{0.973, 0.259, 0.012}
\definecolor{pitch_a}{rgb}{0.965, 0.816, 0.035}
\definecolor{pitch_b}{rgb}{0.835, 0.961, 0.000}
\definecolor{pitch_fis}{rgb}{0.247, 0.976, 0.008}
\definecolor{pitch_cis}{rgb}{0.024, 0.969, 0.659}
\definecolor{pitch_gis}{rgb}{0.024, 0.439, 0.969}

\definecolor{label_a}{HTML}{fad4d3}
\definecolor{label_b}{HTML}{ffaaaa}
\definecolor{label_c}{HTML}{ffff00}
\definecolor{label_d}{HTML}{c1ddb7}
\definecolor{label_e}{HTML}{a7c2ab}
\definecolor{label_f}{HTML}{c0c0ff}
\definecolor{label_g}{HTML}{adedff}
\definecolor{label_h}{HTML}{e1c9e7}

\newtcbox{\rectLabelA}{enhanced,nobeforeafter,tcbox raise base,boxrule=\boxRule,top=\boxTopLine,bottom=\boxBottomLine,
  right=\boxRightLine,left=\boxLeftLine,arc=\boxArc,boxsep=\boxSep,before upper={\vphantom{dlg}},
  colframe=label_a,coltext=black,colback=label_a}
\robustify{\rectLabelA}
\newtcbox{\rectLabelB}{enhanced,nobeforeafter,tcbox raise base,boxrule=\boxRule,top=\boxTopLine,bottom=\boxBottomLine,
  right=\boxRightLine,left=\boxLeftLine,arc=\boxArc,boxsep=\boxSep,before upper={\vphantom{dlg}},
  colframe=label_b,coltext=black,colback=label_b}
\robustify{\rectLabelB}
\newtcbox{\rectLabelC}{enhanced,nobeforeafter,tcbox raise base,boxrule=\boxRule,top=\boxTopLine,bottom=\boxBottomLine,
  right=\boxRightLine,left=\boxLeftLine,arc=\boxArc,boxsep=\boxSep,before upper={\vphantom{dlg}},
  colframe=label_c,coltext=black,colback=label_c}
\robustify{\rectLabelC}
\newtcbox{\rectLabelD}{enhanced,nobeforeafter,tcbox raise base,boxrule=\boxRule,top=\boxTopLine,bottom=\boxBottomLine,
  right=\boxRightLine,left=\boxLeftLine,arc=\boxArc,boxsep=\boxSep,before upper={\vphantom{dlg}},
  colframe=label_d,coltext=black,colback=label_d}
\robustify{\rectLabelD}
\newtcbox{\rectLabelE}{enhanced,nobeforeafter,tcbox raise base,boxrule=\boxRule,top=\boxTopLine,bottom=\boxBottomLine,
  right=\boxRightLine,left=\boxLeftLine,arc=\boxArc,boxsep=\boxSep,before upper={\vphantom{dlg}},
  colframe=label_e,coltext=black,colback=label_e}
\robustify{\rectLabelE}
\newtcbox{\rectLabelF}{enhanced,nobeforeafter,tcbox raise base,boxrule=\boxRule,top=\boxTopLine,bottom=\boxBottomLine,
  right=\boxRightLine,left=\boxLeftLine,arc=\boxArc,boxsep=\boxSep,before upper={\vphantom{dlg}},
  colframe=label_f,coltext=black,colback=label_f}
\robustify{\rectLabelF}
\newtcbox{\rectLabelG}{enhanced,nobeforeafter,tcbox raise base,boxrule=\boxRule,top=\boxTopLine,bottom=\boxBottomLine,
  right=\boxRightLine,left=\boxLeftLine,arc=\boxArc,boxsep=\boxSep,before upper={\vphantom{dlg}},
  colframe=label_g,coltext=black,colback=label_g}
\robustify{\rectLabelG}
\newtcbox{\rectLabelH}{enhanced,nobeforeafter,tcbox raise base,boxrule=\boxRule,top=\boxTopLine,bottom=\boxBottomLine,
  right=\boxRightLine,left=\boxLeftLine,arc=\boxArc,boxsep=\boxSep,before upper={\vphantom{dlg}},
  colframe=label_h,coltext=black,colback=label_h}
\robustify{\rectLabelH}
\newcommand{\vertOne}{\docircle{white}{black}{black}{1}}
\newcommand{\vertTwo}{\docircle{white}{black}{black}{2}}
\newcommand{\vertThree}{\docircle{white}{black}{black}{3}}
\newcommand{\vertFour}{\docircle{white}{black}{black}{4}}
\newcommand{\vertNine}{\docircle{white}{black}{black}{9}}

\newcommand{\aLabel}{\rectLabelA{A}}
\newcommand{\bLabel}{\rectLabelB{B}}
\newcommand{\cLabel}{\rectLabelC{C}}
\newcommand{\dLabel}{\rectLabelD{D}}
\newcommand{\eLabel}{\rectLabelE{E}}
\newcommand{\fLabel}{\rectLabelF{F}}
\newcommand{\gLabel}{\rectLabelG{G}}
\newcommand{\hLabel}{\rectLabelH{H}}

\newtcbox{\rectBlkWhtBtn}{enhanced,nobeforeafter,tcbox raise base,boxrule=\boxRule,top=\boxTopLine,bottom=\boxBottomLine,
  right=\boxRightLine,left=\boxLeftLine,arc=\boxArc,boxsep=\boxSep,before upper={\vphantom{dlg}},
  colframe=black,coltext=black,colback=white}
\robustify{\rectBlkWhtBtn}
\definecolor{stroke}{rgb}{0.2, 0.2, 0.2}

\newtcbox{\rectC}{enhanced,nobeforeafter,tcbox raise base,boxrule=\boxRule,top=\boxTopLine,bottom=\boxBottomLine,
  right=\boxRightLine,left=\boxLeftLine,arc=\boxArc,boxsep=\boxSep,before upper={\vphantom{dlg}},
  colframe=pitch_c,coltext=black,colback=white}
\newtcbox{\rectD}{enhanced,nobeforeafter,tcbox raise base,boxrule=\boxRule,top=\boxTopLine,bottom=\boxBottomLine,
right=\boxRightLine,left=\boxLeftLine,arc=\boxArc,boxsep=\boxSep,before upper={\vphantom{dlg}},
colframe=pitch_d,coltext=black,colback=white}
\newtcbox{\rectE}{enhanced,nobeforeafter,tcbox raise base,boxrule=\boxRule,top=\boxTopLine,bottom=\boxBottomLine,
right=\boxRightLine,left=\boxLeftLine,arc=\boxArc,boxsep=\boxSep,before upper={\vphantom{dlg}},
colframe=pitch_e,coltext=black,colback=white}
\newtcbox{\rectF}{enhanced,nobeforeafter,tcbox raise base,boxrule=\boxRule,top=\boxTopLine,bottom=\boxBottomLine,
right=\boxRightLine,left=\boxLeftLine,arc=\boxArc,boxsep=\boxSep,before upper={\vphantom{dlg}},
colframe=pitch_f,coltext=black,colback=white}
\newtcbox{\rectG}{enhanced,nobeforeafter,tcbox raise base,boxrule=\boxRule,top=\boxTopLine,bottom=\boxBottomLine,
right=\boxRightLine,left=\boxLeftLine,arc=\boxArc,boxsep=\boxSep,before upper={\vphantom{dlg}},
colframe=pitch_g,coltext=black,colback=white}
\newtcbox{\rectA}{enhanced,nobeforeafter,tcbox raise base,boxrule=\boxRule,top=\boxTopLine,bottom=\boxBottomLine,
right=\boxRightLine,left=\boxLeftLine,arc=\boxArc,boxsep=\boxSep,before upper={\vphantom{dlg}},
colframe=pitch_a,coltext=black,colback=white}
\newtcbox{\rectB}{enhanced,nobeforeafter,tcbox raise base,boxrule=\boxRule,top=\boxTopLine,bottom=\boxBottomLine,
right=\boxRightLine,left=\boxLeftLine,arc=\boxArc,boxsep=\boxSep,before upper={\vphantom{dlg}},
colframe=pitch_b,coltext=black,colback=white}
\newtcbox{\rectFis}{enhanced,nobeforeafter,tcbox raise base,boxrule=\boxRule,top=\boxTopLine,bottom=\boxBottomLine,
right=\boxRightLine,left=\boxLeftLine,arc=\boxArc,boxsep=\boxSep,before upper={\vphantom{dlg}},
colframe=pitch_fis,coltext=black,colback=white}
\newtcbox{\rectCis}{enhanced,nobeforeafter,tcbox raise base,boxrule=\boxRule,top=\boxTopLine,bottom=\boxBottomLine,
right=\boxRightLine,left=\boxLeftLine,arc=\boxArc,boxsep=\boxSep,before upper={\vphantom{dlg}},
colframe=pitch_cis,coltext=black,colback=white}
\newtcbox{\rectGis}{enhanced,nobeforeafter,tcbox raise base,boxrule=\boxRule,top=\boxTopLine,bottom=\boxBottomLine,
right=\boxRightLine,left=\boxLeftLine,arc=\boxArc,boxsep=\boxSep,before upper={\vphantom{dlg}},
colframe=pitch_gis,coltext=black,colback=white}

\robustify{\rectC}
\robustify{\rectD}
\robustify{\rectE}
\robustify{\rectF}
\robustify{\rectG}
\robustify{\rectA}
\robustify{\rectB}
\robustify{\rectFis}
\robustify{\rectCis}
\robustify{\rectGis}

%% file: figures.tex
\newcommand{\sldEightQubitsWidth}{0.99\textwidth}
\newcommand{\sldEightQubits}{
\begin{figure}[ht]
    \begin{minipage}{\columnwidth}
        \vspace*{2pt}
        \href{https://graphstatevis.github.io/app?graph=8_0000000}{
            \begin{overpic}[width=\sldEightQubitsWidth,tics=10]{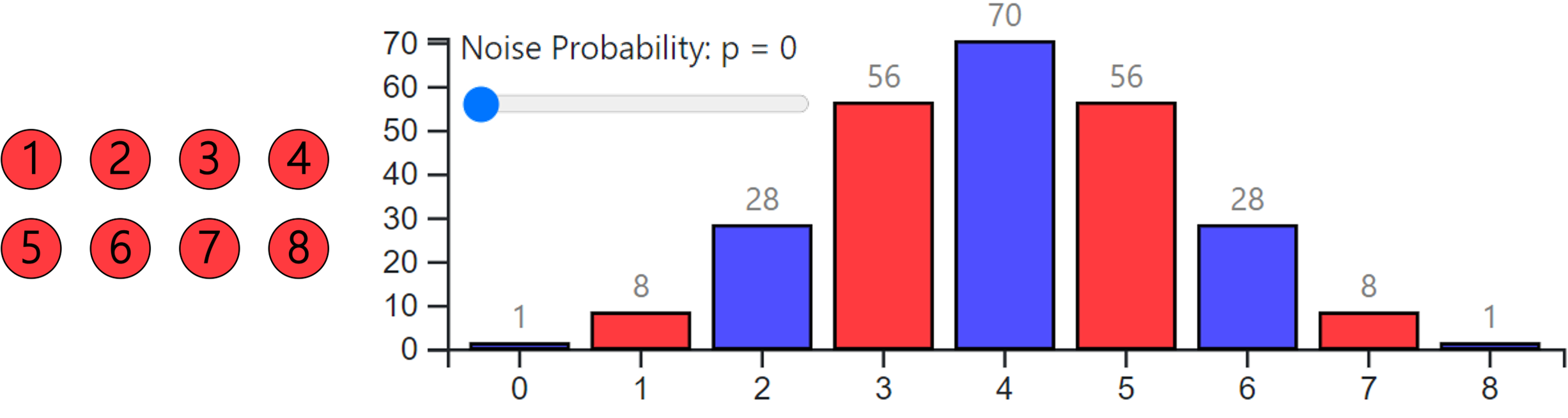}
                \put (0,22) {(a)}
                \put (22.5,26.5) {$A_k$}
                \put (67,-1.5) {$k$}
            \end{overpic}
            \ \\[-7pt]
        }
        \href{https://graphstatevis.github.io/app?graph=8_FE00000}{
            \hspace*{-1pt}\begin{overpic}[width=\sldEightQubitsWidth,tics=10]{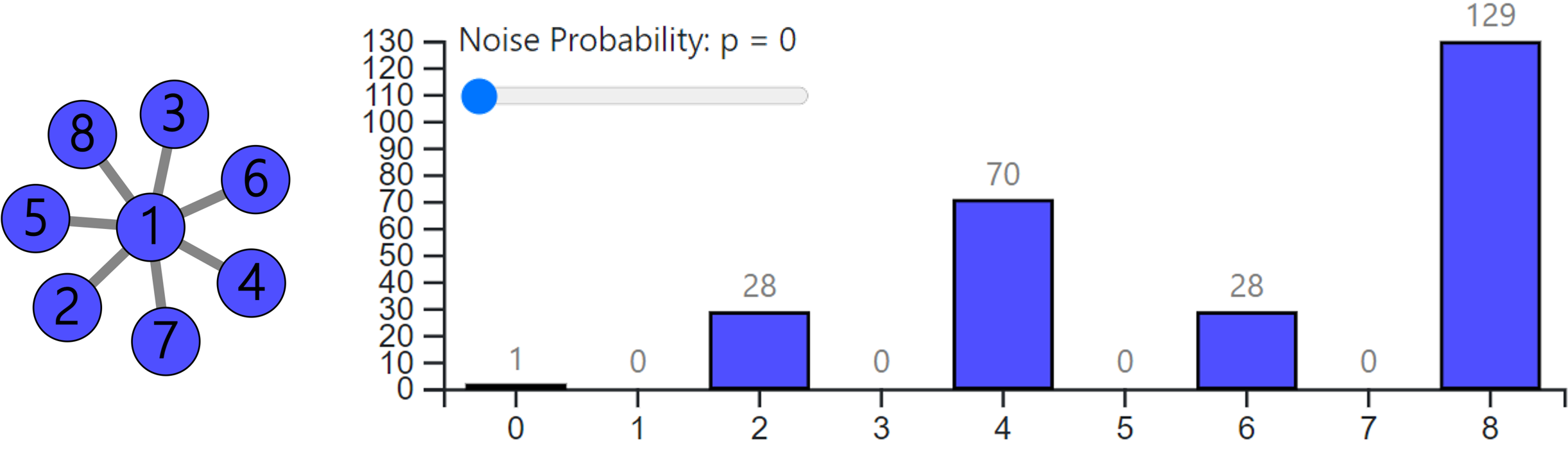}
                \put (0,25) {(b)}
                \put (22.5,29.5) {$A_k$}
                \put (67,-1.5) {$k$}
            \end{overpic}
            \ \\[-7pt]
        }
        \href{https://graphstatevis.github.io/app?graph=8_9124865}{
            \hspace*{-1pt}\begin{overpic}[width=\sldEightQubitsWidth,tics=10]{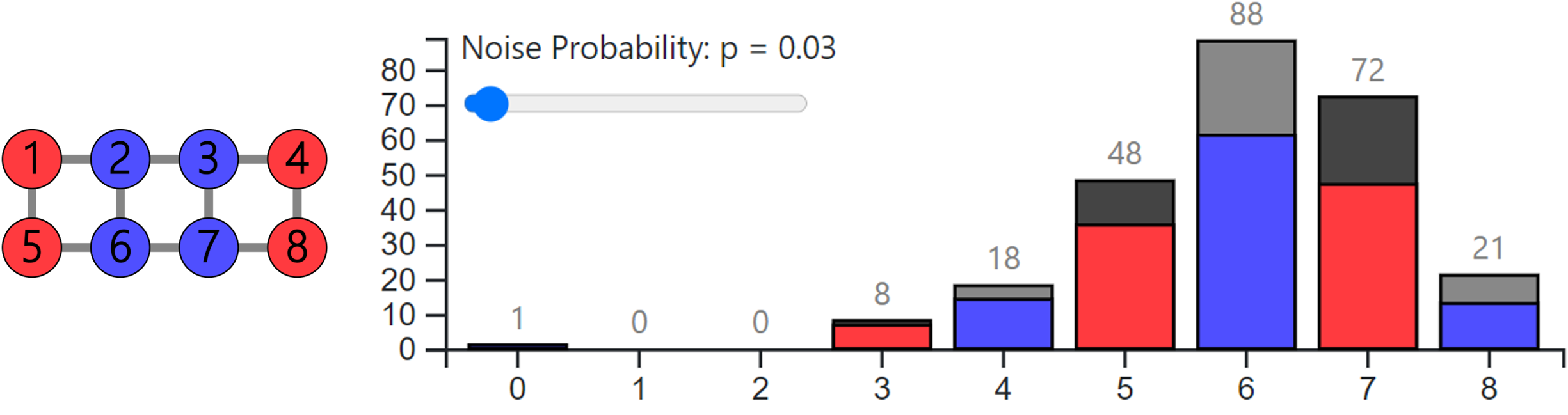}
                \put (0,22) {(c)}
                \put (22.5,26) {$A_k$}
                \put (67,-1.5) {$k$}
            \end{overpic}
            \vspace*{-5pt}
        }
    \end{minipage}
    \caption{Examples of eight-vertex graphs with activated \emph{Parity Coloring} (left) and their corresponding SLD histograms (right). 
    The bar height $A_k$ displays the frequency of weight-$k$  operators in the stabilizer group of the corresponding graph state.
    The SLDs in panel (a) and (c) are of \emph{type~I} because the corresponding graphs contain a (red) vertex with an even number of neighbors, while the SLD in panel (b) is of \emph{type~II} because there are only (blue) vertices with an odd number of neighbors.
    When the noise parameter $p$ (slider) is increased to a nonzero value,
    the partially decayed SLD of the noisy state $\mathcal{E}_{p}^{\otimes n}[\ket{G}\bra{G}]$ is displayed in red and blue, while a gray-scale SLD histogram of the noise-free state remains visible in the background
    (c). 
    }
    \label{fig:eight_qubit_slds}
\end{figure}
}

\newcommand{\sldSydney}{
    \begin{figure*}[htb] 
        \centering
        \href{https://graphstatevis.github.io/app?graph=27_400000140000080000040000040000100002000008000500000000400020014004001000801010020A0020A5}{
            \hspace*{-4pt}\begin{overpic}[width=0.99\textwidth,tics=10]{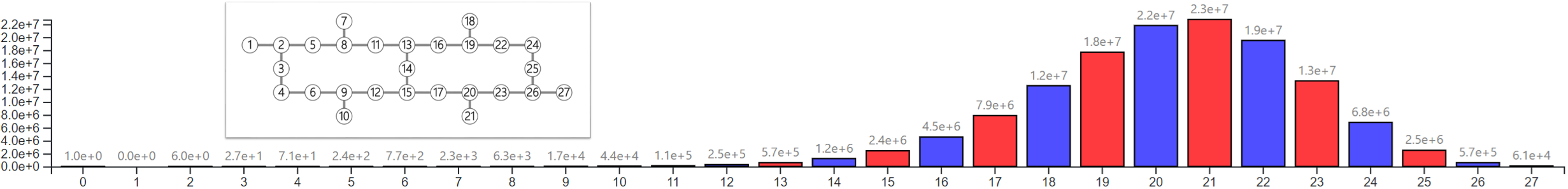}
                \put (5,9.5) {(a)}
                \put (0.5,12) {$A_k$}
                \put (100.5,0.5) {$k$}
            \end{overpic}
            \vspace*{10pt}
        }
        \href{https://graphstatevis.github.io/app?graph=27_00000014000008000004000004000010000000000800010000000040002000400000100080100002020020A4}{
            \hspace*{-4pt}\begin{overpic}[width=0.99\textwidth,tics=10]{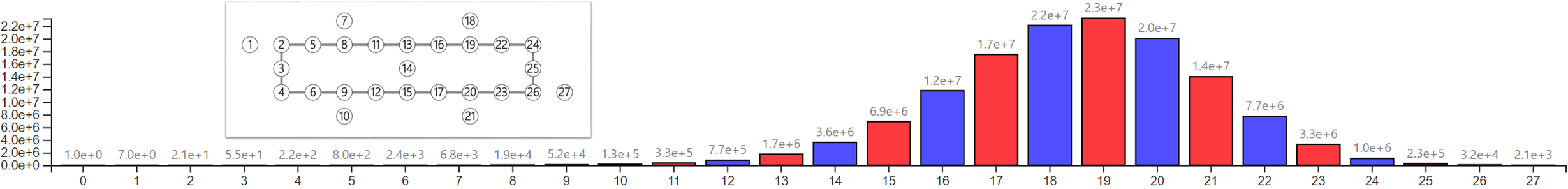}
                \put (5,9) {(b)}
                \put (0.5,11.75) {$A_k$}
                \put (100.5,0.5) {$k$}
            \end{overpic}
        }
        \vspace*{-15pt}
        \caption{Graphs (inset) and SLDs of hardware efficient graph states.
        The graph in panel (a) and (b), respectively, coincides with the connectivity graph of \ibmqSydney{} and one of its subgraphs.
        The SLD is equal to the Pauli-weight distribution of the operators, which can be simultaneously measured by uncomputing the graph state, followed by a qubit-wise readout.
        }
        \label{fig:sld27}
    \end{figure*}
}

\newcommand{\sldGHZlike}{
    \begin{figure*}[htb] 
        \centering
        
        \href{https://graphstatevis.github.io/app?graph=27_7FFFFFE000000000000000000000000000000000000000000000000000000000000000000000000000000000}{
        \hspace*{-7pt}\begin{overpic}[width=0.99\textwidth,tics=10]{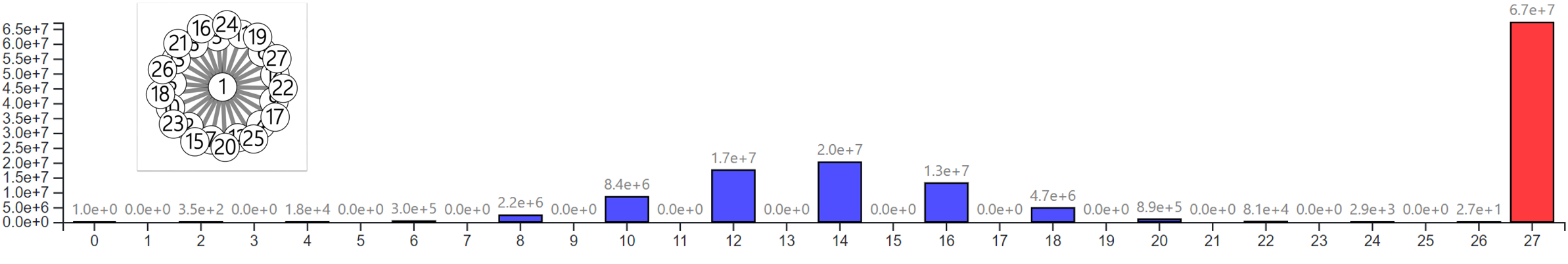}
            \put (5,14) {(a)}
            \put (1,16.5) {$A_k$}
            \put (100.5,1.5) {$k$} 
        \end{overpic}
        }
        \vspace*{2pt}
        \href{https://graphstatevis.github.io/app?graph=27_7FFFFF8000000000000000000000000000000000000000000000000000000000000000000000000000000000}{
        \hspace*{-3pt}\begin{overpic}[width=0.99\textwidth,tics=10]{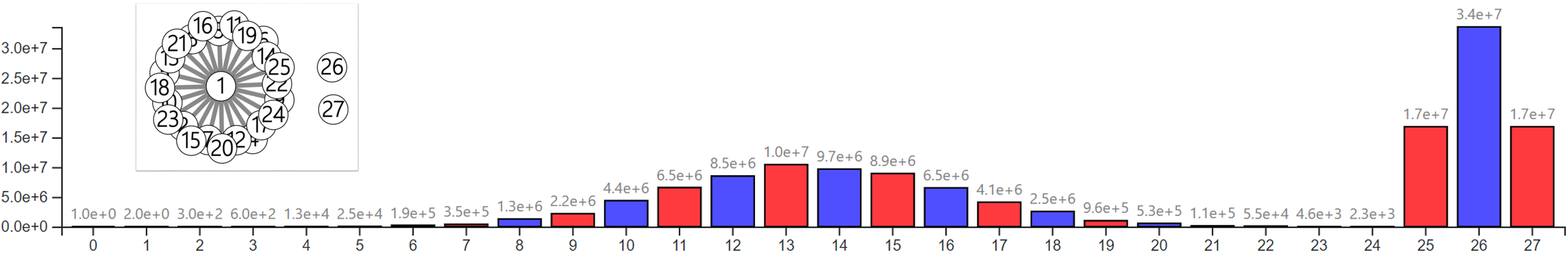}
            \put (5,13.75) {(b)}
            \put (1,16) {$A_k$}
            \put (100.5,1) {$k$} 
        \end{overpic}
        }
        \vspace*{2pt}
        \href{https://graphstatevis.github.io/app?graph=27_7FFFFE0000000000000000000000000000000000000000000000000000000000000000000000000000000000}{
        \hspace*{-3pt}\begin{overpic}[width=0.99\textwidth,tics=10]{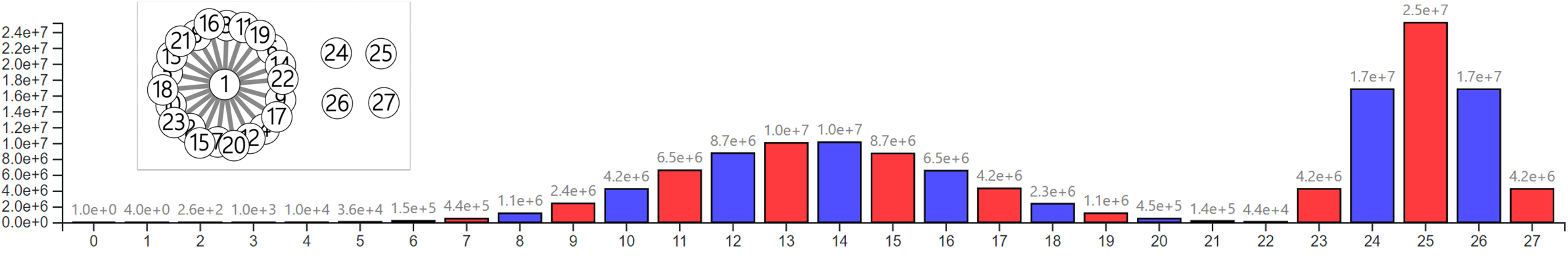}
                \put (5,14) {(c)}
                \put (1,16) {$A_k$}
                \put (100.5,1) {$k$}
            \end{overpic}
        }
        \vspace*{-5pt}
        \caption{Graph sum of an $m$-vertex star graph and $n-m$ isolated vertices (inset) and SLDs of their graph states (see \autoref{eq:SLD_GHZ_and_product}), where $n=27$ and $m=27,25,$ and $23$ for panel (a), (b), and (c), respectively.
        The SLDs feature a sharp peak at $k=n-m/2$, which could help tailoring VQE measurements.
        }
        \label{fig:sld27_ghz}
    \end{figure*}
}

%% file: 0_abstract.tex
\begin{abstract}

Fathoming out quantum state space is a challenging endeavor due to its exponentially growing dimensionality.
At the expense of being bound in its expressiveness, the discrete and finite subspace of graph states is easier to investigate 
via a pictorial framework accompanied with a theoretical toolkit from the stabilizer formalism.
Analyzing hand-drawn graphs is a tedious and time-consuming task and imposes limitations to the problem sizes that can be addressed.
Similarly, algorithmic studies using adjacency matrices alone lack the benefit of a visual representation of the states.
We argue that applying visual analytics to investigate graph states can be advantageous.
To this end, we introduce \GraphStateVis, a web-based application for the visual analysis of qubit graph states and their stabilizer groups.
Our tool facilitates the interactive construction of a graph through multiple components supported by linking and brushing.
The user can explore graph-state-specific properties, including the Pauli-weight distribution of its stabilizer operators and noise thresholds for entanglement criteria.
We propose a use case in the context of near-term quantum algorithms to illustrate the capabilities of our prototype. 
We provide access to \GraphStateVis{} as an open-source project and invite the broader quantum computing and engineering communities to take advantage of this tool and further boost its development.

\end{abstract}

%% file: 1_Introduction.tex
 \section{Introduction}
The advent of quantum computers increasingly stimulates interdisciplinary research and development between computer and quantum information science~\cite{nielsen_quantum_computation_2000, lidar_quantum_error_2013, isakov_understanding_quantum_2016, 
boyd_silicon_chip_2018, tang_a_quantum_2019}.
Visual analytics is a growing subfield of computer science, which integrates the user into the analysis process to better understand complex relationships in data~\cite{sacha_knowledge_2014}. 
For instance, visual analytics has already been successfully applied to investigate graph structures in social networks~\cite{DBLP:conf/iknow/FedericoAMWZ11}.
We argue that the quantum information community can also benefit from using visual analytics  because the state space of a quantum computer is  high-dimensional and complex.

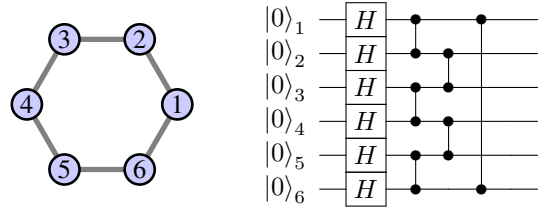
\begin{figure}
    \centering 
    \begin{minipage}{12em}
        \centering
         \scalebox{1}{\input{6RCL_graph}} 
    \end{minipage}
    \begin{minipage}{12em}
        \centering
        \input{6RCL_circuit} 
    \end{minipage}
    \caption{A ring graph with six vertices (left) and the quantum circuit, which prepares the corresponding 6-qubit ring graph state (right). 
    The layer of Hadamard gates prepares every qubit in state $H\ket{0}=\ket{+}$.
    The order of the subsequent two-qubit gates  (one for each edge in the graph) is irrelevant because $\CZ$-gates commute with each other. 
    }
    \label{fig:graph_state_example}
\end{figure}

While a register of $n$ classical bits can be simply described by bit strings $\mathbf{b}=(b_1,\ldots,b_n)\in \FF_2^n$,
an $n$-qubit register can also be in a quantum superposition state $\ket{\psi}= \sum_{\mathbf{b}\in \FF_2^n} z_\mathbf{b} \ket{\mathbf{b}}$.
Hereby, the only constraint on the probability amplitudes $z_\mathbf{b}\in \CC$ is the normalization condition $\sum_{\mathbf{b}} \vert z_{\mathbf{b}}\vert^2=1$.
This flexibility enables novel quantum information protocols having no analog in the classical computing paradigm~\cite{nielsen_quantum_computation_2000}.
When approaching a given problem, quantum information scientists often first investigate \emph{graph states}, which allow for visual illustration~\cite{hein_multiparty_entanglement_2004}.
For an $n$-qubit register, one considers undirected, simple graphs $G=(V,E)$, where $V=\{1,\ldots, n\}$ is the set of vertices. 
Every choice of edge set $E$ leads to a different graph state $\ket{G}$, which is defined as follows.
First, every qubit is initialized in state $\ket{+}= \frac{1}{\sqrt{2}}(\ket{0}+\ket{1})$. 
Then, for every edge $\{i,j\}\in E$ one applies a controlled-$Z$ gate, $\CZ = \diag(1,1,1,-1)$, between qubit $i$ and qubit $j$~(see~\autoref{fig:graph_state_example}).
The resulting state can be written as 
\input{prototype}
\begin{align} \label{eq:graph_state}
    \ket{G} = \frac{1}{\sqrt{2^n}} \sum_{\mathbf{b}\in \FF_2^n} (-1)^{\sum\limits_{j=1}^n\sum\limits_{k=j+1}^n b_j \gamma_{j,k} b_k} \ket{\mathbf{b}},
\end{align}
where $\gamma_{j,k}=1$ if there is an edge $\{j,k\}\in E$, and $\gamma_{j,k}=0$ otherwise. 
The binary matrix $\Gamma=(\gamma_{j,k})_{j,k}\in \FF_2^{n\times n}$ is called \emph{adjacency matrix} of the graph $G=(V,E)$.
%
Graph states play an essential role in the stabilizer formalism~\cite{gottesman_phd_1999}, which relies on the concept of \emph{Pauli operators}
\begin{align}
    X= \begin{pmatrix}
    0&1\\1&0
    \end{pmatrix}, \ 
    Y = \begin{pmatrix} 
    0&-i \\i&0
    \end{pmatrix}, \text{ and }\ 
    Z= \begin{pmatrix}
    1&0 \\ 0&-1
    \end{pmatrix}.
\end{align}
Since $Y = iXZ$, any $n$-qubit Pauli operator can be written as
\begin{align} \label{eq:Pauli}
    i^qX^\mathbf{r}Z^\mathbf{s} = i^q\bigotimes_{k=1}^n X^{r_k}Z^{s_k},
\end{align}
for some bit strings $\mathbf{r},\mathbf{s}\in \FF_2^n$ and an integer $q$ modulo $4$.
A central result of the stabilizer formalism is that, given a group $\mathcal{S}$ of $2^n$ commuting Pauli operators with $-\mathbbm{1}\not\in\mathcal{S}$, there is a unique $n$-qubit state $\ket{\psi}$ with the property $S\ket{\psi}= \ket{\psi}$ for all $S\in\mathcal{S}$.
One calls $\ket{\psi}$ a \emph{stabilizer state} and $\mathcal{S}$ its \emph{stabilizer group}.
Every graph state $\ket{G}$, as defined in Eq.~\eqref{eq:graph_state}, is a stabilizer state  with stabilizer group  $\mathcal{S}=\{ \pm X^\mathbf{r}Z^{\Gamma \mathbf{r}} \ \vert \ \mathbf{r}\in \FF_2^n \}$~\cite{hein_multiparty_entanglement_2004, miller_small_quantum_2019}.
Vice versa,  every stabilizer state $\ket{\psi}$ can be transformed into a graph state $\ket{G}$ by applying suitable single-qubit gates, i.e., $\ket{\psi}$ and $\ket{G}$ are local-unitary (LU) equivalent~\cite{schlingemann_stabilizer_codes_2002}.
Since many quantum state properties do not change under such local transformations, the graph state formalism provides a valuable paradigm for studying quantum entanglement and multipartite correlation in general~\cite{hein_multiparty_entanglement_2004,miller_sector_length_2021}. 
Further applications of graph states contain, but are not limited to,
quantum error correction~\cite{lidar_quantum_error_2013}, simultaneous measurements of commuting Pauli operators~\cite{miller_sector_length_2021}, and measurement-based quantum computation~\cite{raussendorf_measurement_based_2003}.
 
In this paper, we contribute \GraphStateVis, a web-based interactive prototype for visualizing graph states and their stabilizer groups. 
This application enabled us to discover many results in Ref.~\cite{miller_sector_length_2021}.
As we expect our prototype to be stimulating for the broader quantum computing and engineering community, we make it publicly available as an open source project~\cite{graphstatevis2021}.

%

%% file: 6RCL_graph.tex
\begin{tikzpicture}  
\draw[-, line width=.2em, color=gray] (-0.3,0.866) -- (0.3,0.866);  
\draw[-, line width=.2em, color=gray] (-0.3,-0.866) -- (0.3,-0.866);  
\draw[-, line width=.2em, color=gray] (0.9, 0.1732) -- (0.6,0.6928);  
\draw[-, line width=.2em, color=gray] (0.9, -0.1732) -- (0.6,-0.6928);  
\draw[-, line width=.2em, color=gray] (-0.9, 0.1732)  -- (-0.6,0.6928);  
\draw[-, line width=.2em, color=gray] (-0.9, -0.1732) --(-0.6,-0.6928);  
 
\fill[color=blue, opacity=.2] (1,0) circle (0.2);    
\fill[color=blue, opacity=.2] (-1,0) circle (0.2);     
\fill[color=blue, opacity=.2] (-0.5, 0.866) circle (0.2);     
\fill[color=blue, opacity=.2] (-0.5,-0.866) circle (0.2);    
\fill[color=blue, opacity=.2] (0.5, 0.866) circle (0.2);     
\fill[color=blue, opacity=.2] (0.5,-0.866) circle (0.2);   

\draw[line width=.1em] (1,0) circle (0.2);    
\draw[line width=.1em] (-1,0) circle (0.2);     
\draw[line width=.1em] (-0.5, 0.866) circle (0.2);     
\draw[line width=.1em] (-0.5,-0.866) circle (0.2);    
\draw[line width=.1em] (0.5, 0.866) circle (0.2);     
\draw[line width=.1em] (0.5,-0.866) circle (0.2);   

\draw ( 1  , 0    ) node[text=black]{1};
\draw ( 0.5, 0.866) node[text=black]{2};
\draw (-0.5, 0.866) node[text=black]{3};
\draw (-1  , 0    ) node[text=black]{4};
\draw (-0.5,-0.866) node[text=black]{5};
\draw ( 0.5,-0.866) node[text=black]{6};
 \end{tikzpicture}  

%% file: 6RCL_circuit.tex
 
$
    \Qcircuit @C=1.0em @R=0.0em @!R {
	 	\lstick{ {\ket{0}}_{1}  } & \gate{H} & \ctrl{1} & \qw & \ctrl{5} & \qw & \qw\\
	 	\lstick{ {\ket{0}}_{2}  } & \gate{H} & \control\qw & \ctrl{1} & \qw & \qw & \qw\\
	 	\lstick{ {\ket{0}}_{3}  } & \gate{H} & \ctrl{1} & \control\qw & \qw & \qw & \qw\\
	 	\lstick{ {\ket{0}}_{4}  } & \gate{H} & \control\qw & \ctrl{1} & \qw & \qw & \qw\\
	 	\lstick{ {\ket{0}}_{5}  } & \gate{H} & \ctrl{1} & \control\qw & \qw & \qw & \qw\\
	 	\lstick{ {\ket{0}}_{6}  } & \gate{H} & \control \qw & \qw & \control\qw & \qw & \qw\\
	 }
	 $

%% file: prototype.tex
\begin{figure*}[t]
\begin{center}
\href{\appurl}{\includegraphics[width=\textwidth]{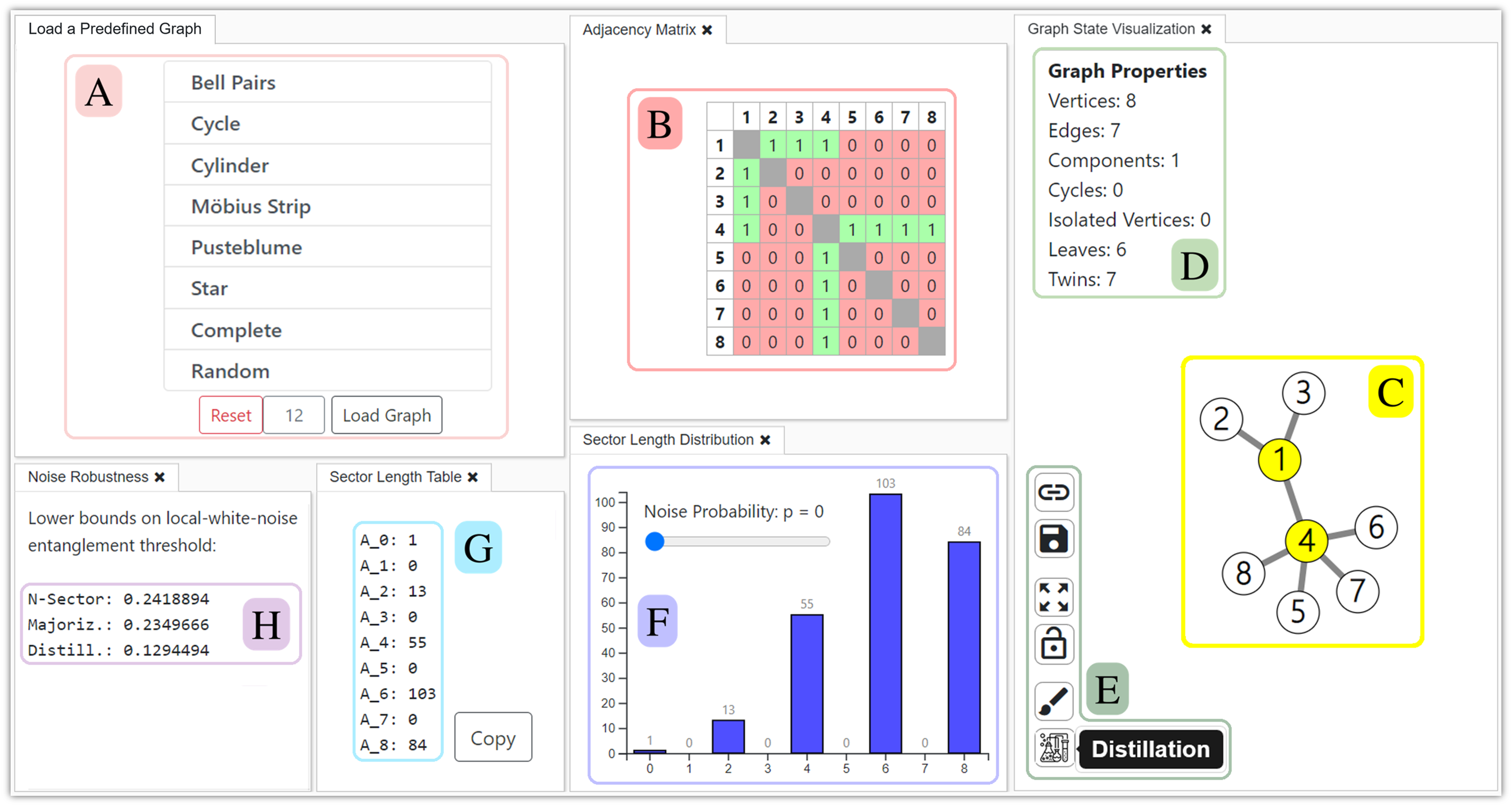}}

\end{center}
\vspace*{-0.3cm}
\caption{\GraphStateVis~is an interactive, web-based application enabling visual exploration and analysis of qubit graph states and their stabilizer groups. 
The application consists of six components.
At the start of the analysis, the user can \emph{Load a Predefined Graph}~\aLabel. 
The \emph{Adjacency Matrix}~$\Gamma$ shows which vertices are connected through edges~\bLabel. 
The \emph{Graph State Visualization} widget displays the current graph~\cLabel~and various \emph{Graph Properties}~\dLabel. 
Several buttons allow the user to export graphs and manipulate the visualization~\eLabel.
The \emph{Sector Length Distribution} (SLD) of the graph state is provided in the form of a histogram~\fLabel~and in the form of a table~\gLabel.
A slider allows visualizing the impact of noise on the SLD histogram (see \autoref{fig:eight_qubit_slds}).
The critical value, below which entanglement is guaranteed by 
specific entanglement criteria,
is provided in the last widget, \emph{Noise Robustness}~\hLabel. 
}
\label{fig:prototype:labeled}
\end{figure*}

%% file: 2_PrototypeDesign.tex
\section{Prototype Design} \label{sec:2}  

The interface of \GraphStateVis{} comprises six integrated components facilitating the exploration and analysis of graph states as illustrated in~\autoref{fig:prototype:labeled}.
Our application allows the user to construct graphs and investigate the corresponding graph states through linking and brushing~\cite{shneiderman_eyes_1996}.

\subsection{Constructing graphs}
 \label{sec:2a}
 

When \GraphStateVis{} is launched, a \emph{Pusteblume} graph with $n=8$ vertices is initialized as a starting graph, which can be modified through mouse interactions in the \emph{Graph State Visualization} widget~\cLabel.
It is also possible to start with a different graph by choosing from a list
in the \emph{Load a Predefined Graph} widget~\aLabel. 
Furthermore, the user can create a \emph{random graph} for which the edge-creation probability $p$ can be  adjusted using a slider, which appears in widget~\aLabel~once this option is selected. 
For example, setting $p=0$ and $p=1$ leads to the \emph{edgeless} and \emph{complete graph}, respectively.
 
\begin{figure}[h]
    \vspace*{-0.1cm}
    \centering
    \includegraphics[width=\linewidth]{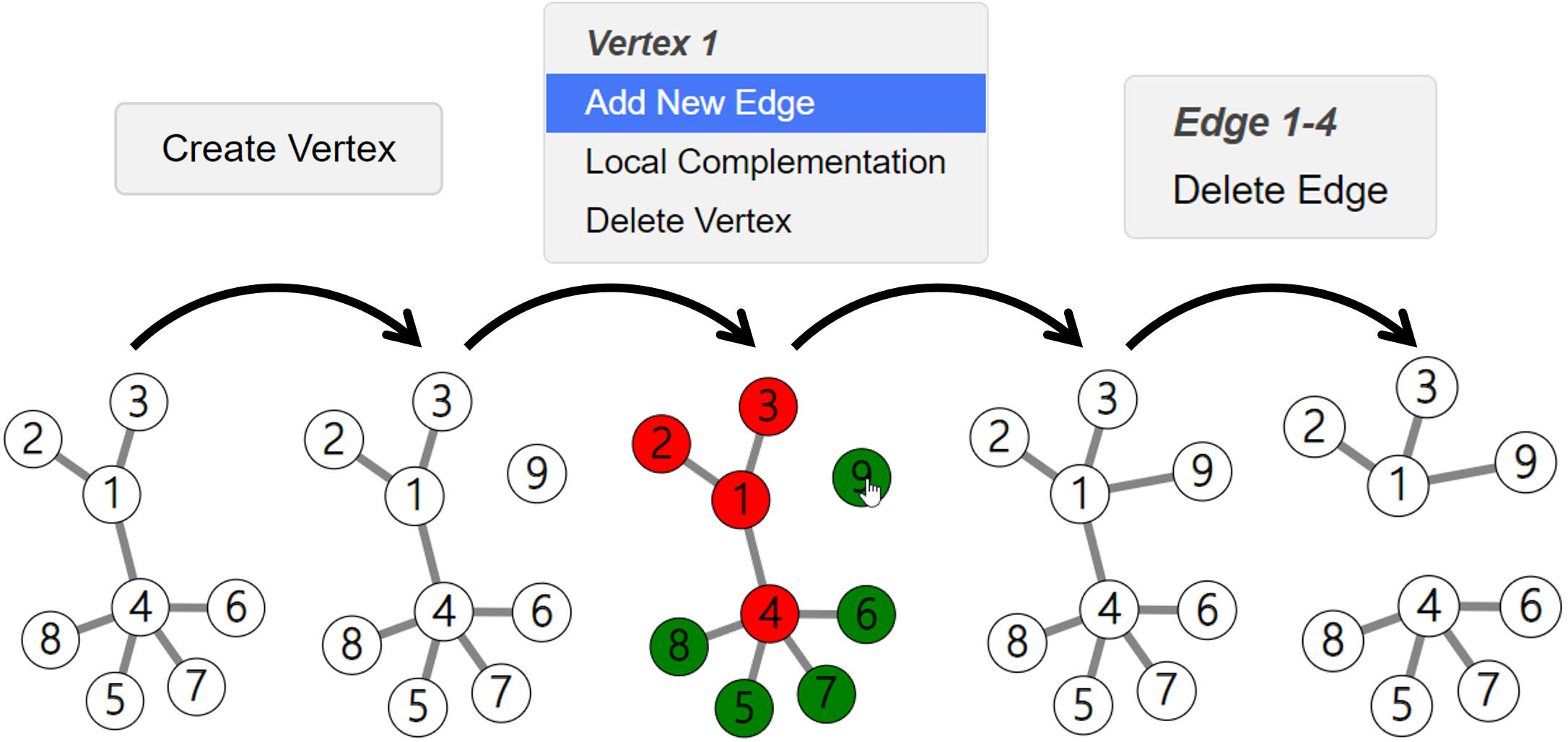}
    \vspace*{-0.35cm}
    \caption{
    Typical workflow of manipulating a graph in our application. 
    A right-click on the background opens a context menu facilitating the \emph{creation} of a new isolated vertex~\vertNine.
    With another right-click on~\vertOne, the user can \emph{add a new edge} to an available (green) vertex; 
    vertices can be unavailable (red) because simple graphs cannot contain loops or multiple edges. 
    Similarly, the edge~\vertOne---\vertFour~can be \emph{deleted} via the edge context menu.}
    \label{fig:workflow}
\end{figure}

To create or remove vertices and edges, the user can follow the graph manipulation workflow as described in~\autoref{fig:workflow}.
Whenever the graph is changed, its adjacency matrix~\bLabel~is updated accordingly. 
Vice versa, it is also possible to create (and delete) edges by flipping the respective entries of the adjacency matrix from 0 to 1 (and back). 
The rows and columns are explicitly enumerated, and a right-click on one of these header cells opens the same vertex context menu as in \autoref{fig:workflow}.
Besides \rectBlkWhtBtn{Add New Edge}~and \rectBlkWhtBtn{Delete Vertex}, there is a third option in this menu: \rectBlkWhtBtn{Local Complementation}~\cite{bouchet_recognizing_locally_1993}. 
Clicking this button will invert the neighborhood of vertex $i$; i.e., for each pair of vertices $j,k$ with $\gamma_{i,j}=\gamma_{i,k}=1$, the value of $\gamma_{j,k}$ is flipped. 
If a sequence of local complementations relates two graphs, then the respective graph states only differ by the application of certain single-qubit Clifford gates; in particular, they are LU equivalent~\cite{hein_multiparty_entanglement_2004}.

By default, the prototype reduces overlapping of vertices, leading to balanced graph representations by exploiting the force-directed graph layout algorithm that is available through the D\textsuperscript{3} library~\cite{d3bostock2011} ~\cLabel.
The automatic layout animation can be toggled by using the \emph{Lock / Unlock Force} button~\forceUnLock~in the \emph{graph state visualization} widget~\eLabel.
In particular, the user can deactivate the repelling forces between vertices to create a manual layout tailored to his or her preference.
%
Triggering the button \emph{Zoom to Fit} \zoomtofit~above will automatically adjust the view to exploit the available space displaying the whole graph.
The user can flexibly navigate the graph visualization via zoom and pan functionality. 

The user can copy the \emph{graph ID}~\copyGraphId~or a specific URL~\copyUrlIcon~into the clipboard to share a configured graph or save it for later reference~\eLabel.
The graph ID consists of the number of vertices followed by the upper half of the adjacency matrix converted into hexadecimal representation. 
This unambiguous mapping allows a graph to be created based on its graph ID, which 
consists of approximately $\frac{n(n-1)}{8}$ characters for a graph with $n$ vertices.
In our GitHub repository~\cite{graphstatevis2021}, we provide python functions that convert a Graph ID into the corresponding adjacency matrix and vice versa.
We expect this graph ID conversion functionality accompanied by the URL loading and saving feature to be beneficial to those researchers who prefer to interactively debug and configure graphs instead of mere coding.

 \subsection{Analyzing the graph state and its stabilizer group}
 \label{sec:2b}
 
 A core functionality of \GraphStateVis{} is to facilitate the investigation of the stabilizer group of a given $n$-qubit graph state $\ket{G}$ with adjacency matrix $\Gamma\in \FF_2^{n\times n}$.
 As it is not feasible to depict all $2^n$ stabilizers individually, the prototype instead displays the sector length distribution \fLabel~(SLD) 
 $\mathbf{A}=(A_0, \ldots, A_n)$
 of the graph state~\cite{aschauer_local_invariants_2004};
 the $k$-body sector length (SL) of a general $n$-qubit state $\rho \in \CC^{2^n\times 2^n}$ is defined as 
 \begin{align}
        A_k [ \rho] = \sum_{\substack{P \in \{\mathbbm 1,X,Y,Z\}^{\otimes n}\\ \wt(P)=k}} 
     \Tr[\rho P]^2 ,
 \end{align}
where  $k\in\{0,\ldots,n\}$ and $\wt(P)$ denotes the weight of a Pauli operator $P$, 
i.e., the number of its non-identity tensor factors. 
 It is well known that for graph states (and, more generally, for stabilizer states), the $k$-body SL coincides with the number of weight-$k$ stabilizer operators $S\in \mathcal{S}=\{(-1)^{\sum_{i<j} r_i  \gamma_{i,j}r_j } X^\mathbf{r}Z^{\Gamma \mathbf{r}} \ \vert\  \mathbf{r}\in \FF_2^n\}$~\cite{kloeckl_characterizing_multipartite_2015}.\footnote{ For example, the three-qubit star-graph state~\scalebox{.9}{\vertOne---\vertTwo---\vertThree}~has one weight-0 stabilizer ($\mathbbm1\mathbbm1\mathbbm1$), no weight-1 stabilizers, three weight-2 stabilizers ($XZ\mathbbm1$, $\mathbbm1ZX$, $X\mathbbm1 X$) and four weight-3 stabilizers ($ZXZ$, $ZYY$, $YYZ$, $-YXY$). Therefore, its SLD is given by $\mathbf{A}=(1,0,3,4)$.}
 Thus, we can write
 \begin{align} \label{eq:SLD_graphstate}
     A_k [ \ket{G}\bra{G}] = \# \{ \mathbf{r} \in \FF_2 ^n \ \vert \ \swt(\mathbf{r},\Gamma \mathbf{r})=k \},
 \end{align}
 where the symplectic weight 
 of a pair of  vectors $\mathbf{r},\mathbf{s}\in~\FF_2^n$ is defined as
  \begin{align}
 \swt(\mathbf{r},\mathbf{s}) = \# \{i\in\{1,\ldots, n\}\ \vert \ r_i=1 \text{ or } s_i=1\}.
 \end{align}
\autoref{eq:SLD_graphstate} gives rise to a brute force algorithm, which we have implemented in \GraphStateVis{} to compute the SLDs of those graph states, which correspond to the connectedness components of the graph built by the user.
These partial SLDs are then recursively combined using the formula
\begin{align} \label{eq:SLD_combination}
    A_k[\rho\otimes \rho'] = \sum_{j=0}^k A_j[\rho]A_{k-j}[\rho'],
\end{align}
which holds for arbitrary multi-qubit states $\rho$ and $\rho'$~\cite{Wyderka_2020}.
The SLD-computation runtime is therefore limited by the size of the largest connected subgraph. It doubles with every new vertex in this subgraph; taking hours for a connected graph with $n\approx 30$ verices.
For this reason, the SLD histogram~\fLabel~is automatically updated only for connected graphs with $n\le 16 $ vertices (and an equivalent for disconnected graphs).
If the SLD is not provided automatically, the user can still enforce its calculation by clicking the button \rectBlkWhtBtn{Compute Sector Length Distribution}, appearing in the SLD widget, if needed.
Whenever the SLD of a new graph is calculated, it is cached; the next time the same graph is created, the SLD is immediately displayed without the need to wait. 
As the cache is stored on a remote server, the user can also access  SLDs that have been previously computed by other users.

To facilitate reading, the SLD histogram bars representing $A_k$ are displayed in blue and red color for even and odd $k$, respectively.
For every stabilizer state, the SLD falls into one of two categories~\cite{huber_some_ulams_2018}: 
If every vertex of the graph has an odd number of neighbors, it holds $A_k = 0$ for all odd $k$ (\emph{type~II}). 
Otherwise, it holds $\sum_{k\text{ odd}} A_k = \sum_{k\text{ even}} A_k$ (\emph{type~I}). 
In \GraphStateVis, an SLD of \emph{type~II} is quickly recognized by viewing only blue bars in component~\fLabel, while for an SLD of \emph{type~I} the blue and red bars cover the same area (see~\autoref{fig:eight_qubit_slds}).
By clicking the button \emph{Parity Coloring}~\parityColoringIcon, every vertex with an even and odd number of neighbors is colored red and blue, respectively~\eLabel.
Thus, the SLD is of \emph{type~II} if and only if all vertices are colored in blue.
This can only happen if $n$ is even.
The parity coloring feature facilitates the identification of a near-minimal number of graph modifications, which can turn an SLD of \emph{type I} into \emph{type II} and vice versa.

\sldEightQubits 
 
 Real quantum hardware is always subject to noise.
 A simple noise model is given by the depolarization channel 
 \begin{align}
     \mathcal{E}_p: \rho \longmapsto p\frac{ \mathbbm1}{2} + (1-p) \rho,
 \end{align}
 which replaces a single-qubit state $\rho$ by a maximally mixed state $\mathbbm 1/2$ with probability $p$~\cite{nielsen_quantum_computation_2000,lidar_quantum_error_2013}.
 The slider in widget~\fLabel~allows the user to adapt $p$ and to visualize how $n$ depolarizing channels (one for each qubit) cause the SLD to decay (see panel (c) of \autoref{fig:eight_qubit_slds}).
The visualization is based on the formula 
\begin{align}
    A_k\left[\mathcal{E}_{p}^{\otimes n}[\rho]\right] = (1-p)^{2k} A_k[\rho],
\end{align}
which is derived in Ref.~\cite{miller_sector_length_2021}.
In the absence of noise, every graph state $\ket{G}$ is entangled, except if $G$ is the edgeless graph.
With gradual increase of noise levels, quantum entanglement is eventually lost.
This phenomenon is known as \emph{sudden death of entanglement}~\cite{yu_sudden_death_2009}.
While the exact threshold value $p_\mathrm{thr}$ at which this transition occurs is difficult to determine, entanglement criteria can provide lower bounds on $p_\mathrm{thr}$. 
For example, every $n$-qubit state $\rho$ with $A_n[\rho]>1$ is entangled~\cite{tran_quantum_entanglement_2015} (but there are also highly entangled states $\rho$ with $A_n[\rho]=0$~\cite{kaszlikowski_quantum_correlation_2008}).
In the noise robustness widget~\hLabel, we refer to this sufficient (but not necessary) entanglement criterion as \texttt{N-Sector} criterion.
Similarly, the value of \texttt{Majoriz.} is a lower bound on $p_\mathrm{thr}$ based on the majorization criterion~\cite{nielsen_separable_states_2001}, which implies that every $n$-qubit state $\rho$ with 
$\sum_k(2k-n)A_k[\rho]  > 0$ is entangled~\cite{miller_sector_length_2021}.
The third and final lower bound on $p_\mathrm{thr}$, which is displayed in the noise robustness widget~\hLabel, is based on an entanglement distillation protocol~\cite{hein_entanglement_properties_2005}.
Its value is given by
\begin{align} \label{eq:loc_thresh_dist}
 p_\texttt{Distill.}  = 1- {2^{-2/ \left(2+ \max\limits_{\{i,j\}\in E} \deg(i)+\deg(j)\right) }},
\end{align} 
where $\deg(i)$ denotes the degree of vertex $i\in V$.
In \GraphStateVis{} it is possible to highlight a pair of neighboring vertices $i,j\in V$, which maximizes the expression $\deg(i)+\deg(j)$, by clicking the \emph{Distillation} button~\distillationIcon~in the graph visualization widget~\eLabel.
This option is activated in \autoref{fig:prototype:labeled}~\cLabel, coloring vertex \scalebox{0.8}{$\vertOne$} and \scalebox{0.8}{$\vertFour$} in yellow.

%% file: 3_UseCase.tex
\section{Use-case proposal: Hardware-efficient measurements of commuting Pauli operators}  \label{sec:3}

\sldSydney

The variational quantum eigensolver (VQE) algorithm is a new quantum algorithm that is hoped will outperform the best classical simulation algorithms in the near future by exploiting noisy intermediate-scale quantum computers~\cite{peruzzo_a_variational_2014, McClean_the_theory_2016}.
The goal of VQE is to minimize the expectation value 
 \begin{align}
     \langle H \rangle _{\boldsymbol{\theta}} =
 \bra{\Psi(\boldsymbol{\theta})}  H \psiTheta
 \end{align} of a problem Hamiltonian $H$ by optimizing parameters $\boldsymbol{\theta}$ in the preparation circuit of a trial quantum state $\ket{\Psi(\boldsymbol \theta)}$.
 To estimate this expectation value, the Hamiltonian is expanded into a linear combination of Pauli operators
 \begin{align}
 H = \sum_{i=1}^m \lambda_i P_i,
 \end{align}
 where $\lambda_i\in \RR$ and $P_i\in\{\mathbbm1, X,Y,Z\}^{\otimes n}$ for all $i\in\{1,\ldots,m\}$.
This allows us to obtain  $\langle H \rangle _{\boldsymbol{\theta}}$  as a linear combination of individually measured expectation values $\langle P_i\rangle _{\boldsymbol \theta}$.
Using simultaneous measurements of commuting Pauli operators, it is possible to reduce the total runtime of the VQE algorithm~\cite{kandala_hardware_efficient_2017, hamamura_efficient_evaluation_2019, gokhale_minimizing_state_2019}.
The partitioning of the set  $\mathcal{M}=\{P_1,\ldots, P_m\}$ into commuting subsets, however, is not unique and comes with a trade-off between the number and depth of  measurement circuits (see Table~I of Ref.~\cite{PhysRevX.10.031064} for an overview of available partitioning methods).
From the stabilizer formalism, it is well known that $2^n$ is the maximum number of commuting $n$-qubit Pauli operators $S$  (up to a global phase). 
After possibly replacing some of these operators by $-S$, the resulting set $\mathcal{S}$ is the stabilizer group of a stabilizer state $\ket{\psi}$.
Since $\{Z^\mathbf{r} \ \vert \ \mathbf{r}\in \FF_2^n\}$ is the stabilizer group of $\ket{0}^{\otimes n}$, we can write $\mathcal{S}=\{UZ^\mathbf{r} U^\dagger \}$, where $U$ is a Clifford circuit fulfilling $\ket{\psi}= U \ket{0}^{\otimes n}$.
For all operators in $\mathcal{M}$, which are of the form $P_i=\pm UZ^\mathbf{r}U^\dagger$, we can simultaneously measure the desired expectation values
\begin{align} 
\langle  P_i \rangle_{\boldsymbol{\theta}}
=  \pm  \bra{ \Psi(\boldsymbol{\theta})}  U
     Z^\mathbf{r}  
U^\dagger \psiTheta
\end{align}
by preparing the VQE trial state $\psiTheta$ on a quantum computer, followed by the application of $U^\dagger$ (we refer to this step as uncomputing the stabilizer state $\ket{\psi}$), followed by a qubit-wise readout in the computational basis.
Note that the latter is nothing else but the simultaneous measurement of  $Z^\mathbf{r}$ for all $\mathbf{r}\in \FF_2^n$.
Since state-of-the-art quantum computers are still very noisy, it can be beneficial to partition  $\mathcal{M}$ into (subsets of) stabilizer groups $\mathcal{S}$, for which the corresponding measurement circuits $U^\dagger$ are tailored to the hardware.
The least noisy measurement circuits consist of a single layer of single-qubit gates. 
They measure the stabilizer group of a product state, which is also known as a \emph{tensor product basis} (TPB) in this context~\cite{kandala_hardware_efficient_2017}.
The Pauli-weight distribution of the operators in a TPB is given by the SLD of a fully separable state, $A_k=\binom{n}{k}$ (see panel (a) of Fig.~\ref{fig:eight_qubit_slds}).
In the full Pauli group, however, there are $4\binom{n}{k}3^{k}$  operators having weight $k$, recall \autoref{eq:Pauli}.
Therefore, the probability of drawing a weight-$k$ operator from the uniform distribution over a TPB and over the full Pauli group, respectively, is given by $p_\mathrm{TPB} = \binom{n}{k}2^{-n}$ and $p_\mathrm{PG}=\binom{n}{k}3^k4^{-n}$.
Although $\mathcal{M}$ does not arise as the sample set of a uniform distribution over the Pauli group (but from encoding schemes such as Jordan–Wigner~\cite{McClean_the_theory_2016}),
a similar mismatch occurs for typical VQE instances~\cite{kandala_hardware_efficient_2017}:
If the operators $P_i\in \mathcal{M}$ are assigned to TPBs, those with a low Pauli weight run out first. The remaining (large-weight) operators are not likely to fit into joint TPBs.
Thus, they require are a large number of additional measurement circuits to complete the partitioning of $\mathcal{M}$ into TPBs.
One step in the direction of overcoming this problem (without sacrificing hardware efficiency) is to complement the layer of single-qubit gates in the measurement circuit of a TPB with a single layer of two-qubit gates~\cite{hamamura_efficient_evaluation_2019}.
Up to LU-equivalence, the TPB-only approach corresponds to the measurement of the edgeless graph's stabilizer group, while the second approach also allows for graphs where every vertex is either isolated or a leaf.

The connectivity graph $\Gcon$ of a quantum computer is the graph whose vertices correspond to qubits, while its edges display to which pairs of qubits physical two-qubit gates can be implemented.
For ion trap quantum computers, $\Gcon$ typically is the complete graph~\cite{wright_benchmarking_11_2019}. 
For super- or semiconductor based quantum computers, however, 
the vertices of $\Gcon$ have a limited degree, causing the number of edges to scale linearly in the number of qubits.
In this case, the preparation and, more importantly (for our purposes), the uncomputation of the graph state, which corresponds to $\Gcon$ or one of its subgraphs, is  a hardware efficient quantum circuit.
As we explained above, uncomputing a graph state facilitates the measurement of its stabilizer group.
By adding single qubit Clifford gates at the beginning of the measurement circuit, one can access a multitude of stabilizer groups without altering the Pauli-weight distribution.
This approach significantly enlarges the number of efficiently-measurable stabilizer groups and includes the  approaches of Refs.~\cite{kandala_hardware_efficient_2017} and~\cite{hamamura_efficient_evaluation_2019} as a special case.
Whether relevant instances of the VQE algorithm can benefit from our idea, however, requires further investigation.

We argue that \GraphStateVis{} can support such an investigation.
As discussed in~\autoref{sec:2}, the user can construct the connectivity graph $\Gcon$ and interactively explore its subgraphs by deleting some of its edges.
Simultaneously, the SLD histogram will display the Pauli-weight distribution of the operators, which can be measured simultaneously by uncomputing the corresponding graph state.
In \autoref{fig:sld27}, we depict the connectivity graph $\Gcon$ of \ibmqSydney\footnote{IBM, the IBM logo, and IBM Q are trademarks or registered trademarks of IBM Corp., in the U.S. and/or other countries.} (a quantum computer with 27 superconducting qubits~\cite{yang_testing_scalable_2021}), one of its subgraphs, and their SLDs.
As it is typically observed for graph states with $I$ isolated vertices, their $k$-body SLs  are approximately given by $A_k \approx 2^n\binom{n}{k}p^k(1-p)^{n-k}$, where $p=(3n-I)/4n$~\cite{miller_sector_length_2021}.
Depending on the Pauli-weight distribution of the operators in $\mathcal{M}$ (which are not yet assigned to a measurement circuit), one should explore graph states with an appropriate value of $I$.
In the special case where $\mathcal{M}$ is the sample set of the unit distribution over the Pauli group, one should start with $I=0$.

 \sldGHZlike 
 Greenberger-Horne-Zeilinger (GHZ) states
$\GHZn = \frac{1}{\sqrt{2}} (\ket{0}^{\otimes n}+\ket{1}^{\otimes n})$
lie in the same LU-class as star- and complete-graph states~\cite{hein_multiparty_entanglement_2004}.
Since the star graph does not match the connectivity of super- or semiconducting quantum computers, GHZ states are most efficiently prepared (and uncomputed) recursively:
Since $\ket{\mathrm{GHZ}^{n-1}}$ is symmetrical under qubit exchange, any qubit can serve as control for a CNOT gate targeting the $n$-th qubit initialized in state $\ket{0}$, i.e.,
\begin{align}
    \mathrm{CNOT}_{i,n} \ket{\mathrm{GHZ}^{n-1}}\otimes \ket{0}= \GHZn
\end{align}
for all $i\in\{0,\ldots,n-1\}$.
In this way, the limited connectivity of a quantum chip is leveraged.
From the theory of SLDs it is known that there is no quantum state having more {weight-$n$} stabilizers than $\GHZn$ and its LU-equivalents~\cite{tran_quantum_entanglement_2015, Eltschka2020maximumnbody}. 
Their $k$-body SLs are given by
\begin{align} \label{eq:SLD_GHZ}
    A_k = \binom{n}{k}\delta_{k,\mathrm{even}} + 2^{n-1}\delta_{k,n},
\end{align}
where $\delta_{k,n}$ is the Kronecker symbol and similarly for $\delta_{k,\mathrm{even}}$~\cite{miller_small_quantum_2019, Eltschka2020maximumnbody}. 
Thus, these states are of high importance if only weight-$n$ operators from $\mathcal{M}$ remain to be assigned to measurement circuits.
Similarly, if mostly large-weight operators remain, one can investigate tensor products of $\ket{\mathrm{GHZ}^{n_i}}$ for different choices of $n_i$ (and their LU-equivalents).
By inserting \autoref{eq:SLD_GHZ} into \autoref{eq:SLD_combination}, we find that the $k$-body SL of $\ket{\mathrm{GHZ}^{m}}\otimes \ket{0}^{\otimes n- m}$ is given by 
\begin{align} \label{eq:SLD_GHZ_and_product}
A_k=2^{m-1}\binom{n-m}{k-m}+ \sum_{j=0}^{\lfloor k/2\rfloor} \binom{m}{2j} \binom{n-m}{k-2j},
\end{align}
where we define $\binom{n}{k}=0$ if $k<0$. 
Besides a broad peak at $k=n/2$, these SLDs have a second, sharp peak at $k=n-m/2$ (see \autoref{fig:sld27_ghz}).
Therefore, these states can serve as starting points to design hardware efficient measurement circuits for large-weight operators, which remain after all the low-weight operators in $\mathcal{M}$ have been partitioned into e.g. TPBs.

These examples illustrate how \GraphStateVis{} can support the design of new VQE measurement strategies. 
When pursuing our use case proposal, researchers can flexibly explore the space of available SLDs using our prototype.
We anticipate that this will help them to identify states whose stabilizer groups match the Pauli operators to be measured.

%% file: 4_Conclusion.tex
\section{Conclusion and Outlook}\label{sec:4}

In this paper, we presented \GraphStateVis, an open-source interactive application for the visual analysis of  graph states, which is ready to use~\cite{graphstatevis2021}. 
%
We illustrated how our prototype could be applied when designing new hardware-efficient measurements for variational quantum algorithms.
In addition, we believe that our application can support the identification of local-unitary invariants of graph states and their meaning in terms of graph invariants.
%
%
We consider our approach to be only a first step in applying visual analytics in the field of quantum information theory.
Therefore, we invite intrigued researchers to tailor and extend the features of \GraphStateVis{} as desired.
For example, it could be modified to support the design of new $k$-uniform states by visually highlighting subgraphs that pose an obstruction to $k$-uniformity~\cite{PhysRevA.69.052330}.
%
%
Moreover, the sector-length-decay visualization could be extended to a larger class of error channels by adding noise probability sliders to analyze their combined effects. 
Directions for major extensions of our tool include the visualization of hypergraph states or qudit graph states~\cite{Rossi_2013, DBLP:journals/tvcg/FischerASSKW21}.
Finally, visual analytics could also be applied to investigate how the sector length distribution of a quantum state changes during a noisy quantum computation. 
Insights about such dynamical effects are essential for the design of error mitigation techniques.
A good starting point is the investigation of Clifford circuits with Pauli noise since theoretical tools are available in this case~\cite{miller_propagation_of_2018}.
 
\section*{Acknowledgments}

The authors are thankful for stimulating discussions with Ivano Tavernelli (IBM Research Europe - Zurich) and Nikolai Wyderka.
Felix Huber suggested the term Pusteblume graph.